# Tailored Graphenic Structures Directly Grown on Titanium Oxide Boost the Interfacial Charge Transfer

Roberto Muñoz[a]*, Carlos Sánchez-Sánchez[a], Pablo Merino[a,b], Elena López-Elvira[a], Carmen Munuera[a], Patricia Gant[a], María F. López[a], Andrés Castellanos-Gómez[a], José Angel Martín-Gago[a], Mar García-Hernández[a]

[a] Materials Science Factory, Instituto de Ciencia de Materiales de Madrid, (ICMM-CSIC), Sor Juana Inés de la Cruz 3, E-28049, Madrid, Spain.
[b] Instituto de Física Fundamental, CSIC, Serrano 121, E28006, Madrid, Spain.

*e-mail: rmunoz@icmm.csic.es; marmar@icmm.csic.es

## Highlights

- Tailored graphenic structures, dots rods and continuous films, were successfully fabricated directly on $TiO_2$ substrates using an extremely clean synthesis protocol by Plasma Assisted Chemical Vapor Deposition.

- Chemical analysis demonstrates undoubtedly the formation of Ti-O-C bonds in the interface of carbon-$TiO_2$ hybrid materials.

- High photocurrent density is generated upon illumination through the carbon-$TiO_2$ contact proving an efficient charge transfer.

## Abstract

The successful application of titanium oxide-graphene hybrids in the fields of photocatalysis, photovoltaics and photodetection strongly depends on the interfacial contact between both materials. The need to provide a good coupling between the enabling conductor and the photoactive phase prompted us to directly grow conducting graphenic structures on $TiO_2$ crystals. We here report on the direct synthesis of tailored graphenic structures by using Plasma Assisted Chemical Vapour Deposition that present a clean junction with the prototypical titanium oxide (110) surface. Chemical analysis of the interface indicates chemical bonding between both materials. Photocurrent measurements under UV light illumination manifest that the charge transfer across the interface is efficient. Moreover, the influence of the synthesis atmosphere, gas precursor ($C_2H_2$) and diluents (Ar, $O_2$), on the interface and on the structure of the as-grown graphenic material is assessed. The inclusion of $O_2$ promotes vertical growth of partially oxidized carbon nanodots/rods with controllable height and density. The deposition with Ar results in continuous graphenic films with low resistivity ($\rho=6.8 \cdot 10^{-6}$ $\Omega \cdot m$). The synthesis protocols developed here are suitable to produce tailored carbon-semiconductor structures on a variety of practical substrates as thin films, pillars or nanoparticles.

## 1. Introduction

Since the discovery of the photo-activity of $TiO_2$ under UV light irradiation [1, 2] significant efforts have been made to develop strategies that enhance its "photonic efficiency" or light-to-current conversion capability. The efficiency is usually hindered by the high recombination rate (>90%) of electrons and holes inside the material [3-6] before their diffusion to the surface, where the generated charges can be extracted or expected to promote chemical (redox) reactions [7, 8].

Stimulated by the pioneering work of Kamat and co-workers [9], several attempts have been made to combine graphene or other carbon materials with $TiO_2$ to obtain platforms that minimize the recombination rate [10, 11]. It´s worth mentioning that recent reports strongly manifest that the use of graphene in the nanoscale is in essence the same as using other carbon (nanotubes or activated carbon) materials on enhancement of photoactivity of $TiO_2$[12]. Recently, graphene and graphene oxide mesoscopic flakes with intrinsic defects were successfully used as doping agents to the mesoporous $TiO_2$ electron transport layer in perovskite solar cells, boosting the power conversion efficiency of the device [13]. The use of graphene or carbon structures in contact with the semiconductor under light exposure results in an efficient separation of the yielded charge carriers across the interface of both materials [14-17]. The combination of carbon materials with $TiO_2$ also improves its molecular adsorption capabilities and widens its light absorption range, therefore, enhancing the overall performance [8, 18-20].

A high quality interface between carbon and $TiO_2$ free of recombination centers, such as adsorbates or contamination, and with a well established chemical contact between both materials is critical for charge separation [16, 20]. Importantly, the properties of the interface are strongly depending on the synthetic route [21, 22]. Recently, chemical vapor deposition (in atmospheric pressure and with thermal activation) has been used to directly deposit contamination-free graphitic carbon coatings either on $TiO_2$ single crystals, as a model system [7], and on $TiO_2$ nanoparticles [23] that may present a direct application in photocatalysis [24]. However, the initial stages of carbon nucleation and growth direction of the carbon phase on the $TiO_2$ substrate are aspects not assessed in these works. In close relation to this, a recent theoretical report [25] depicted a scenario in which vertical arrangement of mesoscopic graphene sheets on the substrate may enhance charge separation properties. This work emphasizes the need of "vertical" or columnar growth with interfacial chemical bonds for more efficient charge transfer processes in comparison with graphene-TiO$_2$ flat heterostructures (often referred to as "van der Waals") with clear interface and physical contact.

Tuning the growth mode from its initial states can be achieved by plasma assisted chemical vapor deposition as a versatile technique to control the growth direction of graphene. Thus, horizontal coatings [26, 27] or vertical nanowalls, and nanorods [28] on different insulators and semiconductors have been grown. Moreover, the plasma assisted deposition enhances the degree of activation of the precursor molecules and the substrate surface at lower temperature than thermal synthesis, which positively influences the resulting interfacial bonding.

We here report, for the first time, on a new plasma-based approach for the direct synthesis of graphitic carbon structures, namely atomically thin films, nanodots and nanorods, on rutile TiO$_2$ (110) single-crystal surface. We have chosen this face due to the high stability of this surface [29] and because it is the most common interfacial termination in nanostructured TiO$_2$ (e.g. nanoparticles) providing us a model system to investigate the carbon-TiO$_2$ hybrid. The influence of the deposition atmosphere on the structure, the junction and the growth direction of the graphitic material is assessed. The chemical analysis of the interface indicates that covalent bonds are established between both materials. Subsequent photocurrent measurements under light illumination reveal that the charge transfer across the interface is efficient. The methodologies shown hereafter led to the fabrication of clean carbon-semiconductor interfaces and can be extended to a variety of practical semiconducting structures as thin films, barrier layers[30], pillars or nanoparticles [23, 31].

## 2. Experimental

*Carbon materials synthesis:* Electron Cyclotron Resonance chemical vapor deposition (ECR-CVD) plasma-assisted technique is used for the deposition of graphenic materials. The employed ASTEX AX 4500 system consists of a microwave power source, a two zone chamber and a two stage pumping system [32]. The reaction chamber is polarized. The bias electrodes attract electrons and ions from the plasma generated in the plasma chamber as, in this way, only neutral species reach the sample surface and the formation of defects is greatly minimized. Rutile $TiO_2$ (110), rutile single crystals (Crystal GmbH, ref. TIO 205E, 5x5x0.5 mm$^3$, single side polished) are used as substrates and $C_2H_2$ as carbon precursor diluted in Ar or $O_2$. The surface roughness of these crystals measured by AFM is rms=0.6 Å (fig. S2).The growth process is separated into two steps, with different temperature to control the nucleation (675ºC) and growth stages (735ºC) [26, 33, 34]. In the first step, high quality graphitic seeds nucleate. In the second step, the growth is promoted from the nucleated dots up to the graphenic structure or layer formation. The plasma power (200 W) and precursor ($C_2H_2$) pressure and flow are kept constant throughout the study (see text).

*Characterization:* Room temperature atomic force microscopy (AFM) measurements are performed with a commercial instrument and software from Nanotec [35]. Two different operation modes are employed: dynamic mode, exciting the tip at its resonance frequency ($\sim$75 kHz) to acquire topographic information of the samples and contact mode to sweep away the carbon deposits and measure the thickness. Charge transfer through carbon-$TiO_2$ interface under light illumination is assessed by measuring current vs bias voltage characteristics with a Keithley 2450 source-meter. The photocurrent measurements are performed by the direct microprobing method using carbon fiber tips to avoid the scratching of the surface and minimize the contact potentials with graphene. The probes are made out polyacrylonitrile (PAN) carbon fibers ~1mm long and 7 μm diameter [36]. We illuminate the sample in the UV-Vis range using a Bentham TLS120Xe tunable light source with 8 mW/cm$^2$ constant power density. To modulate the light power we use a variable attenuator in between two optical fibers (one connected to the lamp and the other connected to the set-up) in order to adjust the power of the light at the end of the last optical fiber. The photocurrent is extracted from the measured current under illumination by subtracting the current values in dark conditions. The structure of our composites is addressed by Raman spectroscopy using a confocal Raman microscope (Witec alpha-300R). Raman spectra have been obtained using a 532 nm excitation laser and a 100x objective lens (NA = 0.9). The incident laser power is 1 mW. The nature of the chemical bonds between carbon and $TiO_2$ is determined with X-ray Photoelectron Spectroscopy (XPS). XPS measurements are carried out under ultra-high vacuum (UHV) conditions using a PHOIBOS 100 1D delay line detector electron/ion analyzer, monochromatic Al Kα anode (1486.6 eV), and a pass energy of 15 eV. The binding energy (BE) scale is calibrated with respect to the Ti 2p core level peak at 459.3 eV [37]. All peaks shown in this work are fitted using Voigt functions after subtraction of a Shirley-type background. In all cases, the Lorentzian full-width half-maximum (FWHM-L) is kept constant during the fitting (0.25 and 0.35 eV for C 1s and O 1s, respectively) while the Gaussian one (FWHM-G) is allowed to change. The sheet resistance of the continuous films is characterized by four point probe measurements (JANDEL RMS2 Universal Probe) with continuous current (from 1 to 100 μA). Light absorption is measured by ultraviolet-visible spectroscopy (UV-Vis) by using a SHIMADZU SolidSpec–3700 Spectrophotometer equipped with an integrating sphere.

## 3. Results and discussion

### 3.1 Growth and morphology of graphitic structures on TiO$_2$ (110)

We have used various gas mixtures of O$_2$/C$_2$H$_2$ and Ar/C$_2$H$_2$ for the synthesis rendering a variety of graphenic materials. Figure 1(a) compares the optimized deposition profiles in both cases, with only O$_2$ during synthesis (see blue dashed line) and with only Ar during synthesis (see black dotted line) as diluents. In both protocols, the initial heating step is carried out under an O$_2$ atmosphere to preserve the oxide surface from possible thermal-induced O desorption during this stage. The final cooling step is also carried out under an Ar atmosphere to avoid carbon oxidation. Best results are achieved when 675ºC and 735ºC are used during the nucleation and growth steps, respectively. Processing time, plasma power, partial pressures of O$_2$, Ar and C$_2$H$_2$ are kept constant during the synthesis (see the footnote under fig. 1 for further information of parameters).

Figures 1(b$_1$) and (b$_2$) show the AFM topographic images of the carbon structures grown with an O$_2$/C$_2$H$_2$ atmosphere. It can be clearly observed in the AFM line profile of fig. 1(b$_3$) that a columnar growth occurs, resulting in carbon dots/rods with apparent lateral size of around 50-70 nm and few nm (8-12 nm) in height. The nucleation density is low and well controlled with an average value of 5 nuclei/μm$^2$. The carbon dots are homogeneously distributed on the TiO$_2$ surface, 0.5 cm in lateral size (our plasma deposition system has been previously checked to deposit materials homogeneously in samples up to two cm in diameter). We relate this growth mode to the preservation effect of O$_2$ on the surface. As it can be noticed in Fig. 1(b$_1$-b$_2$), the surface remains flat,

similar to pristine sample, confirming the role of oxygen neutralizing the thermal-induced O desorption from the TiO₂ surface, decreasing its degree of activation and minimizing the nucleation points.

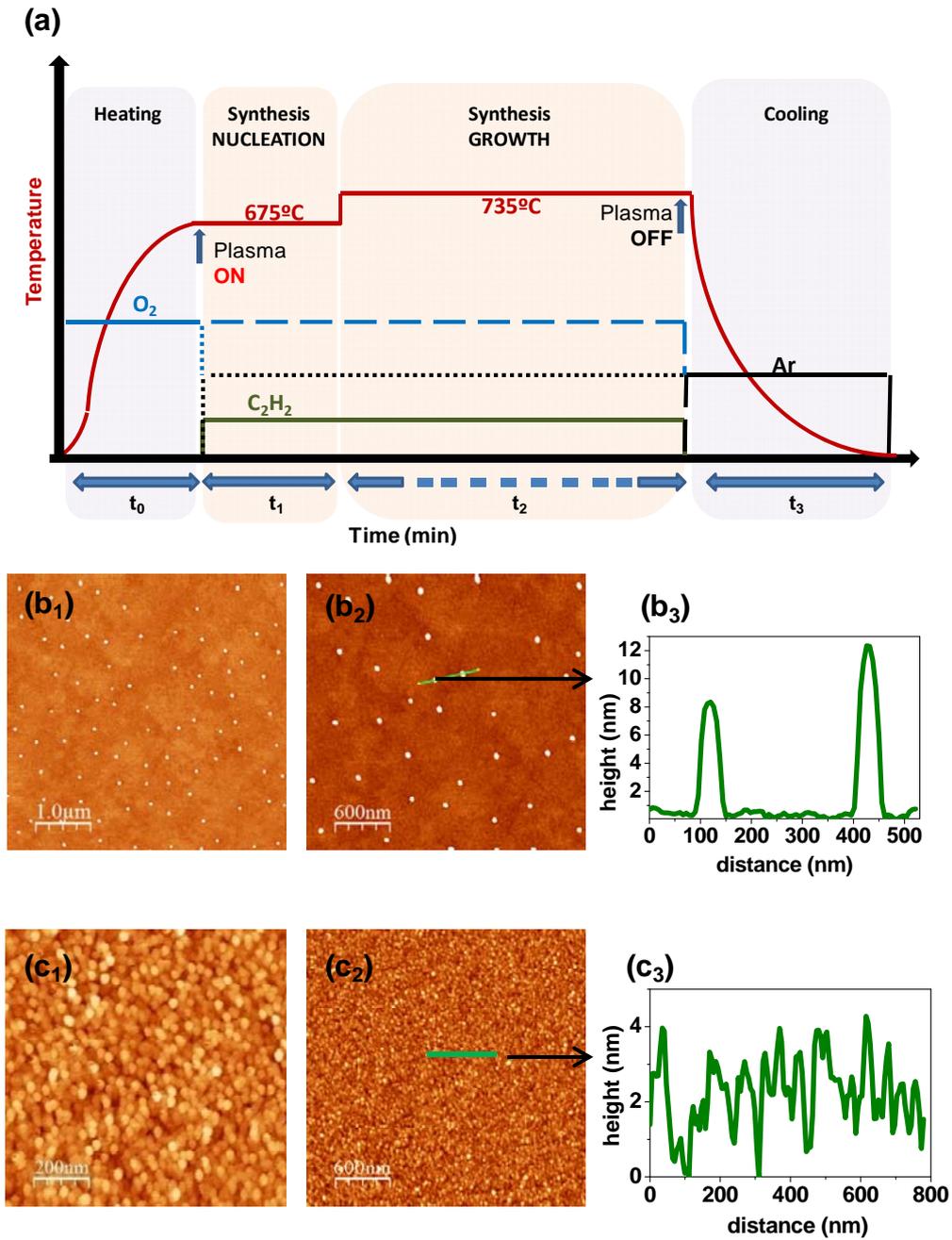

**Figure 1.** (a) Two-step synthesis temperature and partial pressures profiles for carbon dots-rods with O₂ diluent (in blue, dashed line) and continuous films with Ar diluent (in black dotted line). Temperature, processing time and C₂H₂ flow is similar in both cases. Temperature: 675ºC (nucleation), 735ºC (growth), $t_0$=40 min, $t_1$=4 min, $t_2$=12 min, $t_3$=40 min. O₂ flow= 15 sccm, Ar flow= 30 sccm, C₂H₂ flow=2 sccm. Plasma power 200W. (b₁, b2) AFM topography image of carbon rods grown in an O₂/C₂H₂ atmosphere. (b₃) Line profile of two carbon rods from image (b₂) Vertical scale: 0-15 nm. (c₁, c₂) AFM topographic image of a graphitic film deposited in an Ar/C₂H₂ atmosphere. Vertical scale: 0-5 nm. (c₃) Corresponding roughness profile of (c₂).

This resulting graphenic structure deposited over the semiconductor could be suitable for photocatalytic applications as it resembles the desired structure of metal cluster loaded semiconductor photocatalyst [38], even more taking into account that de nucleation density can be controlled, as we show below. Figures 1($c_1$) and ($c_2$) show the AFM images of the deposit performed with an Ar/$C_2H_2$ atmosphere. It can be noticed that in this case the surface is fully covered with a continuous film (see fig. 1($c_2$)) with homogeneous structure and average grain size of 30-50 nm (see the structure in fig. 1($c_1$)) that corresponds to a nucleation density of 400 nuclei/$\mu m^2$, approximately.

Figure 1($c_3$) shows the roughness profile of the film. The film thickness is around 4 nm (see Fig. S1). It is worth mentioning that it is possible to control the thickness of the film with the deposition time even at submonolayer regime. Figure S1 in SI shows that the substrate surface does not suffer any etching during growth. The electrical continuity of this film has been tested by four point probe measurements with an average sheet resistance of 3 k$\Omega \cdot sq^{-1}$ (corresponding $\rho$=6.8$\cdot 10^{-6}$ $\Omega \cdot m$). This high conductivity is related to the graphenic character of the film. Taking into account that the probe separation is 3 mm and film thickness is around 4 nm the continuity of the film accounts for the homogeneous distribution of the material.

The optimized nucleation density of both growth protocols can be compared by observing the AFM images in figs. 1($b_2$) and 1($c_2$). We can speculate that the low nucleation density in the first case is due to the critical role that $O_2$ gas plays in replacing the vast majority of the desorbed terminal O atoms which would promote the generation of vacancies on the surface due to the high temperature during growth [29]. The balance between O removal and O replacement affects the nucleation rate and strongly depends on the diluents used. In the case of Ar, balance is not achieved which results in a surface activation that promotes a noticeable amount of nucleation centers. The activation of the surface is related to its slight reduction and partial functionalization by H atoms generated by the plasma activation of $C_2H_2$. However, this effect is greatly minimized by the fast deposition rate.

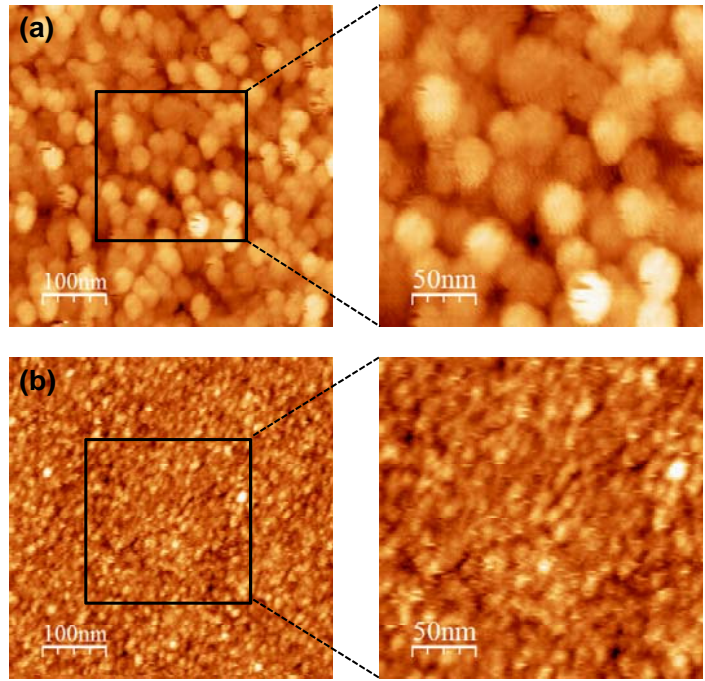

**Figure 2.** (a) AFM topographic images of the continuous film deposited using a two-step methodology with Ar diluent. Temperature: 675ºC (nucleation), 735ºC (growth), $t_0$=40 min, $t_1$=4 min, $t_2$=12 min, $t_3$=40 min. Ar flow= 30 sccm, $C_2H_2$ flow=15 sccm. Plasma power 200W. (b) AFM topographic images of the deposit performed in one step with Ar diluent. Temperature: 735ºC (growth), $t_0$=40 min, $t_1$=0 min, $t_2$=12 min, $t_3$=40 min. Ar flow= 30 sccm, $C_2H_2$ flow=15 sccm. Plasma power 200W. Vertical scale in both figures: 0-5 nm.

In order to further demonstrate the benefits of using a two-step synthesis protocol, figure 2 presents a comparison between the sample grown with a two-step protocol with Ar/$C_2H_2$ and one sample grown with a single-step protocol without the nucleation step and similar growth conditions. Figure 2(a) shows the continuous film structure with an average grain size of 50 nm in diameter (see the amplified image on the right). Figure 2(b) shows the structure of the deposit performed using a one-step growth procedure. In the latter case, it can be clearly observed that the average grain size is less than 20 nm, much smaller than in the two-step case. This smaller grain size has a direct impact in the final resistance of the film (tens of kΩ). It is also important to underline that the lateral increase of the grain observed with the two-step Ar/$C_2H_2$ protocol does not occur in the $O_2/C_2H_2$ protocol. Figure 3 shows that in this case, only an increase in the height of the nuclei and in the nucleation density is observed. Both the nucleation density and the nuclei height can be tuned by modifying the deposition parameters, but the diameter of the grains remains nearly constant. This is the main feature that corroborates that the vertical growth of the nuclei is favored with respect to its horizontal growth.

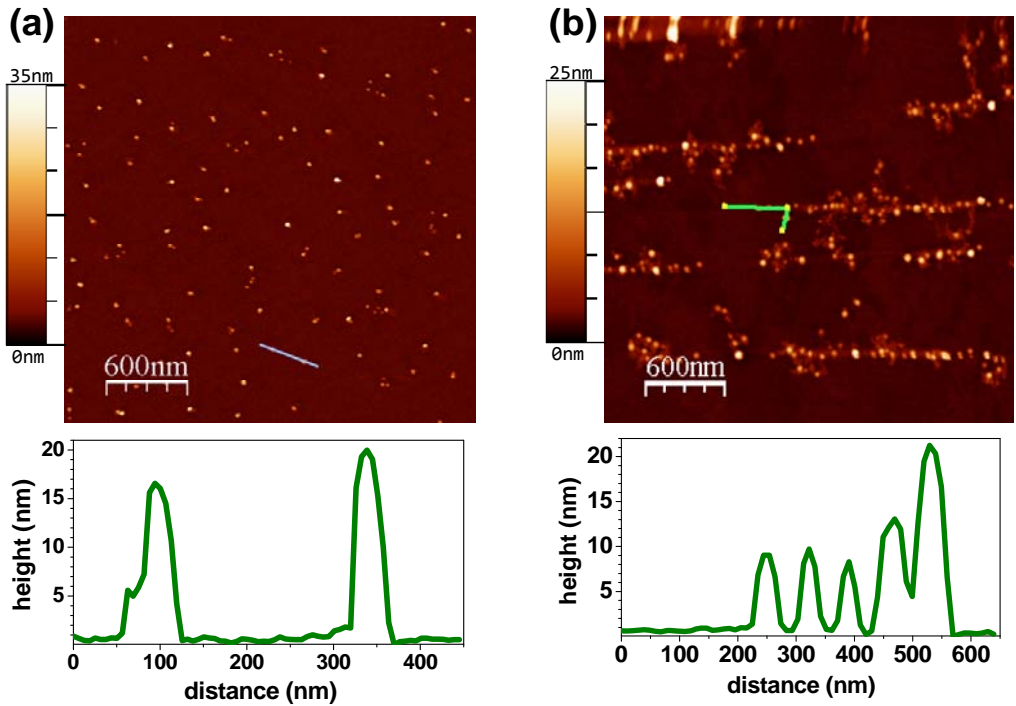

**Figure 3.** AFM topography images of carbon rods deposited with $O_2$ diluent. $t_0$=40 min, $t_1$=4 min, $t_3$=40 min. $O_2$ flow= 15 sccm, $C_2H_2$ flow=2 sccm. Plasma power 200W. (a) Increasing growth time $t_2$=30 min. Temperature: 675ºC (nucleation), 735ºC (growth). (b) Increasing temperature: 695ºC (nucleation), 755ºC (growth), $t_2$=12 min. The corresponding line profiles are below the images.

### 3.2 Structural and surface analysis of the deposits and the interface with $TiO_2$ (110)

Figure 4(a) shows the Raman spectra of the samples deposited in two steps using $Ar/C_2H_2$ (in black) and $O_2/C_2H_2$ (in blue) atmospheres. The Raman spectra from the $TiO_2$ reference crystal (in purple) and sample deposited in one step with $Ar/C_2H_2$ (see fig 2(b)) are also included (in red) for comparison. Both spectra obtained from the samples deposited with Ar present similar structure of the grown material (the intensity is lower in one-step probably related to smaller thickness) and show the typical D, G and 2D carbon peaks, confirming that the deposits are crystalline and graphenic in nature [39, 40].

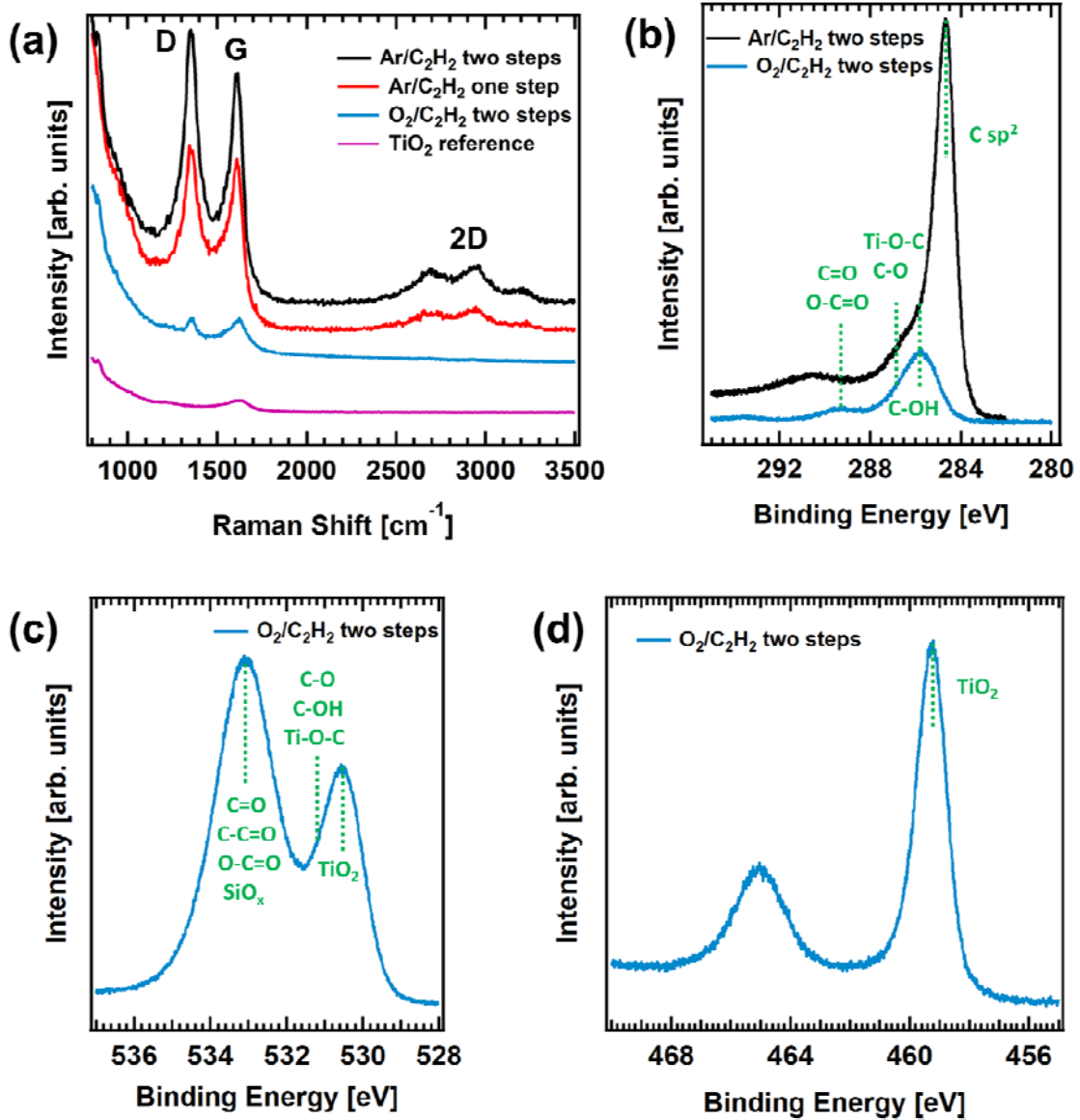

**Figure 4.** (a) Confocal Raman spectra of samples deposited in two steps with $O_2$ diluent (blue line), Ar diluent (black line) and $TiO_2$ reference crystal (purple). The spectrum from the one-step deposition with Ar is also included (red line). (b) High resolution XPS C 1s core level spectra of the samples deposited in two steps with $O_2$ diluent (blue line) and Ar diluent (black line). (c, d) High resolution XPS O 1s and Ti 2p core level spectra of the sample deposited with $O_2$ in two steps showing the main contributions.

The low intensity of the 2D peak, the low $I_{2D}/I_G$ ratio and the high value of the $I_D/I_G$ ratio ($\approx 1$) indicate a small grain size and a high density of grain boundaries, in good agreement with Fig. 2. Along with the grain boundaries, another contribution to the high D peak intensity could be some remaining internal defects within the grains as well as some functionalization with plasma species (i.e. certain amount of hydrogen from the $C_2H_2$ plasma, inducing some $sp^3$ hybridization in the corresponding carbon atoms) [41]. In order to assess more in depth the crystallinity of the continuous film deposited with Ar we subjected the sample to X-Ray diffraction (XRD) measurements (Fig. S3). In principle, the XRD analysis suggests that our sample lacks long range graphitic order. However, the analysis confirms that it is composed by turbostratic graphenic carbon nanoflakes. The spectrum in blue in Figure 4(a), corresponding to the sample deposited with $O_2$ in two steps, shows a D carbon related peak with low intensity, that can be ascribed to the small amount of deposit (5 nuclei/$\mu m^2$) as observed in fig. 1(b) together with a high contribution of the substrate. The carbon G peak (1600 cm$^{-1}$) overlaps with a substrate band (purple line), which is located at a similar shift around 1630 cm$^{-1}$, and 2D peak does not appear in the spectrum. However, the small D peak observed confirms the carbon deposition. It should be noted that the appearance of a clear and relatively narrow D peak (FWHM$\approx 45$ cm$^{-1}$) is related to the tiny size of the dots observed and verifies that the deposit is not amorphous but of crystalline nature [39].

In order to further elucidate the chemical structure and the quality of the interface between the deposits and the $TiO_2$ (110), we perform a surface analysis of the samples with XPS. Figure 4(b) shows the characteristic XPS C 1s core level spectra obtained for both samples, i.e., the one grown with $O_2$ diluent (blue curve) and the sample deposited with Ar diluent (black curve). Comparison between both spectra gives important insights into the chemical composition of both systems. First, we notice the lower intensity of the carbon signal of the sample grown with $O_2$ compared to the sample grown using Ar as diluent; this is a further corroboration of the lower amount of graphenic material obtained using the former approach. Second, the spectrum of the sample grown with $O_2$, shows a wider peak which is shifted toward higher binding energy (BE) with respect to the sample grown with Ar.

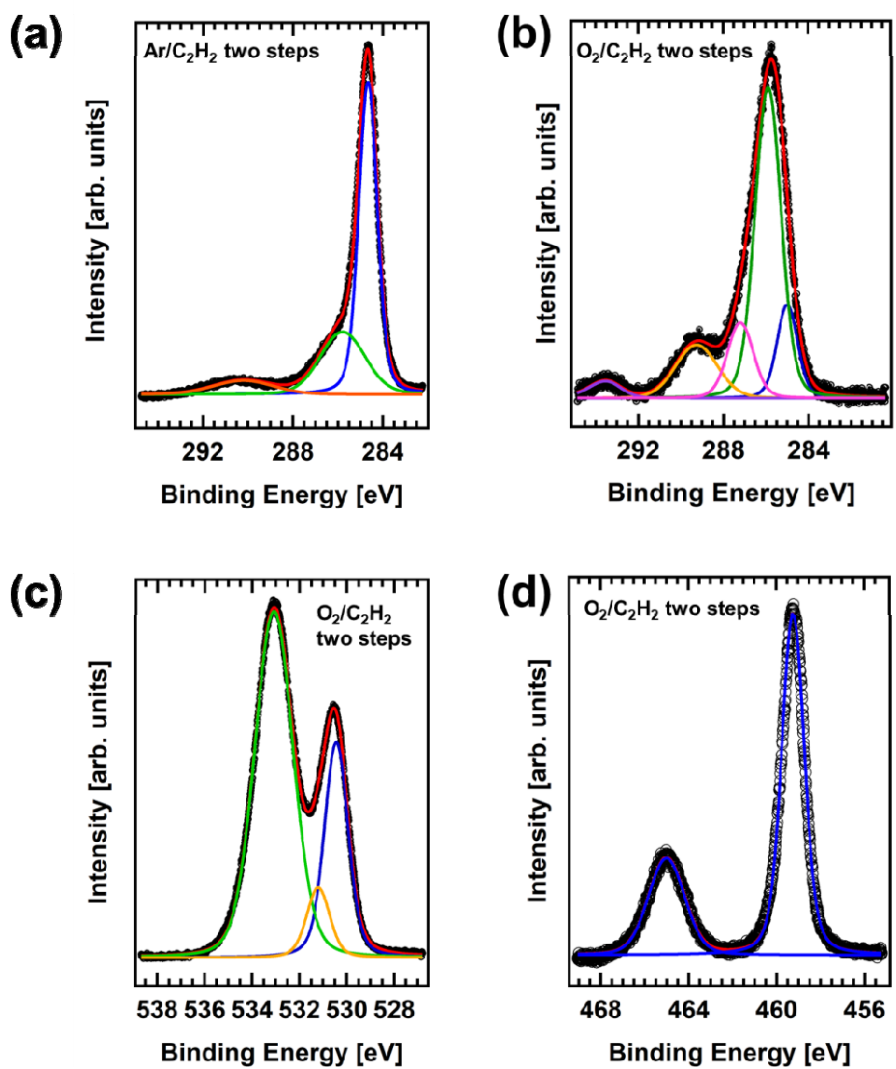

**Figure 5**. Deconvolution of the high resolution XPS peaks. (a) C1s core level peak of the two-steps Ar/C₂H₂ sample. Three different components have been used in the fit. Black dots: raw data; red curve: fit; blue, green, and orange curves: components. (b) C1s core level peak of the two-steps O₂/C₂H₂ sample. Five different components needed to properly fit the data. Black dots: raw data; red curve: fit; blue, green, pink, orange, and violet curves: components. (c) O1s core level peak of the two-steps O₂/C₂H₂ sample. The peak can be decomposed into three different components. Black dots: raw data; red curve: fit; blue, green, and orange curves: components. (d) Ti2p core level peak of the two-steps O₂/C₂H₂ sample. The peak can be fitted with a single doublet component. Black dots: raw data; red curve: fit; blue curve: component.

In order to better understand the origin of the difference in shape and position of the two peaks, we have deconvoluted them into their components (see Fig. 5 (a-b)). The C1s core level associated to the Ar sample is composed of three peaks located at 284.7, 285.8, and 290.2 eV. The first one is characteristic of a sp² configuration, thus compatible with the presence of graphitic C species [42]. The second peak is associated with C-OH species, which could be due to functionalization of grain boundaries. Finally, the third component, presenting a very low area and appearing at 290.2 eV, is assigned to C=O or O-C=O associated to other adsorbed organic species. On the other hand, the deconvolution of the C1s core level peak of the sample prepared under an $O_2$-rich environment in figure 5(b) yields important differences (see also the XPS analysis of the $O_2$ reference sample experiencing similar process without $C_2H_2$ and only with $O_2$, in Fig S4). First of all, it is composed of five components instead of three, with a small component at 285.0 eV corresponding to graphitic $sp^2$ carbon (this contribution is from adventitious carbon in Fig. S4 (a)). This result implies that there is a lower amount of purely $sp^2$ graphitic structures as one would expect. Instead, we observe the presence of a component at 285.9 eV, similar to that found in the other sample, and associated to C-OH, and another component 287.2 eV, associated with C-O, (i.e. Ti-O-C bonds). These two are the majority species (the C-O bond at 287.2 eV is undetectable in reference sample). Then, there are two other minority species appearing at 289.3 and 293.5 eV, that we assign to C=O or O-C=O species, and second order events (i.e. plasmon losses or shake-up features), respectively. Thus, considering all the components present in the $O_2$-rich sample, we can unambiguously state that a great amount of C atoms on the surface are covalently bound to O, thus confirming the formation of oxidized graphene or highly reduced graphene oxide [43-45]. In order to further corroborate the presence of Ti-O-C bonds in the interface, we analyse O1s and Ti 2p peaks from the sample grown with $O_2$ as diluent (see fig. 4(c-d)) and deconvolute them into their components (see fig. 5(c-d) and fig. S4 (b-c)). Figures 4(c) and 5(c) show that three main components can be identified in O1s peak, one at 530.4 eV corresponding to oxygen in $TiO_2$, and another at 531.2 eV, assigned to C-O, C-OH and Ti-O-C bonds [42, 43, 46]. Finally, a third contribution found at around 533.1 eV which is usually attributed to C=O, O-C=O species [42, 47]. However, it can be clearly observed that in this case this component is quite strong. We analyzed in depth the survey and detailed spectra acquired from this sample and from the $O_2$ reference sample experiencing similar process without $C_2H_2$ (see fig. S4 (b) and fig. S5) and we confirmed that there also exists a contribution from impurities to this peak (i.e. SiOx residuals from $TiO_2$ (110) surface polishing procedure). Figure 4(d) shows the XPS Ti2p spectrum for the $O_2$ sample. This spectrum is formed by a single component located at 459.3 eV, in perfect agreement with the expected position for titanium in $TiO_2$ substrates [48]. Other contributions due to titanium carbide formation (Ti-C bond at 455 eV) or substrate reduction ($Ti^{3+}$ at 458 eV) are not detected (see fig. 5(d)). With these results, we propose that the signals observed in the C 1s spectrum at 287.2 eV and O 1s spectrum at 531.2 eV are contributed by Ti-O-C bonds to a large extent. However, as the O1s component at 531.3 eV is present in the $O_2$ reference sample in Fig. S4 (b), the results are not conclusive (it should be noted that, even though this component is present in both samples, we perceive that its ratio with the component from oxygen in $TiO_2$ is higher). Moreover, Ti-O-C component is missing in the spectra acquired from the sample deposited with Ar in two-step. In this case, the signal comes mostly from the continuous graphenic layer (few nm thick) being the signal from the interface weak.

To ascertain the composition of the interface of the deposits made with Ar, we analyse the structure and interface on the sample deposited in one-step as this sample is thinner than the sample grown in two steps and the interfacial contribution to the spectra is expected to be much higher. We also performed the XPS analysis of an Ar reference sample experiencing similar process without $C_2H_2$ and only with Ar. We show in Figure 6, the characteristic XPS C 1s, O 1s and Ti 2p core level spectra obtained from both samples.

Overall, the C1s and O1s core level spectra in fig. 6 (a-d) safely confirm the formation of chemical Ti-O-C bonds between C and O (TiC bond is not detected). The C1s component at 287.2 eV (pink) in Fig. 6(a), that correspond to Ti-O-C bonds, does not appear in the C1s core level peak of the Ar reference sample in Fig. 6(b). Neither it appears in Fig. 5 (a) related to the C1s core level associated to the Ar sample deposited in two-steps. Also, the O1s at 531.7 eV (green), assigned to C-O, C-OH and Ti-O-C bonds is very low in the Ar reference sample in comparison with the deposited sample. This bonding in an oxygen-free atmosphere, where the only source of O is the sample itself, undoubtedly confirms a direct contact between the graphitic film and the $TiO_2$ surface, a crucial characteristic in efficient charge transfer and extraction processes [49]. The formation of this chemical bond has been previously proposed in graphene-$TiO_2$ composites [23, 24, 46, 50, 51] without such clear evidence.

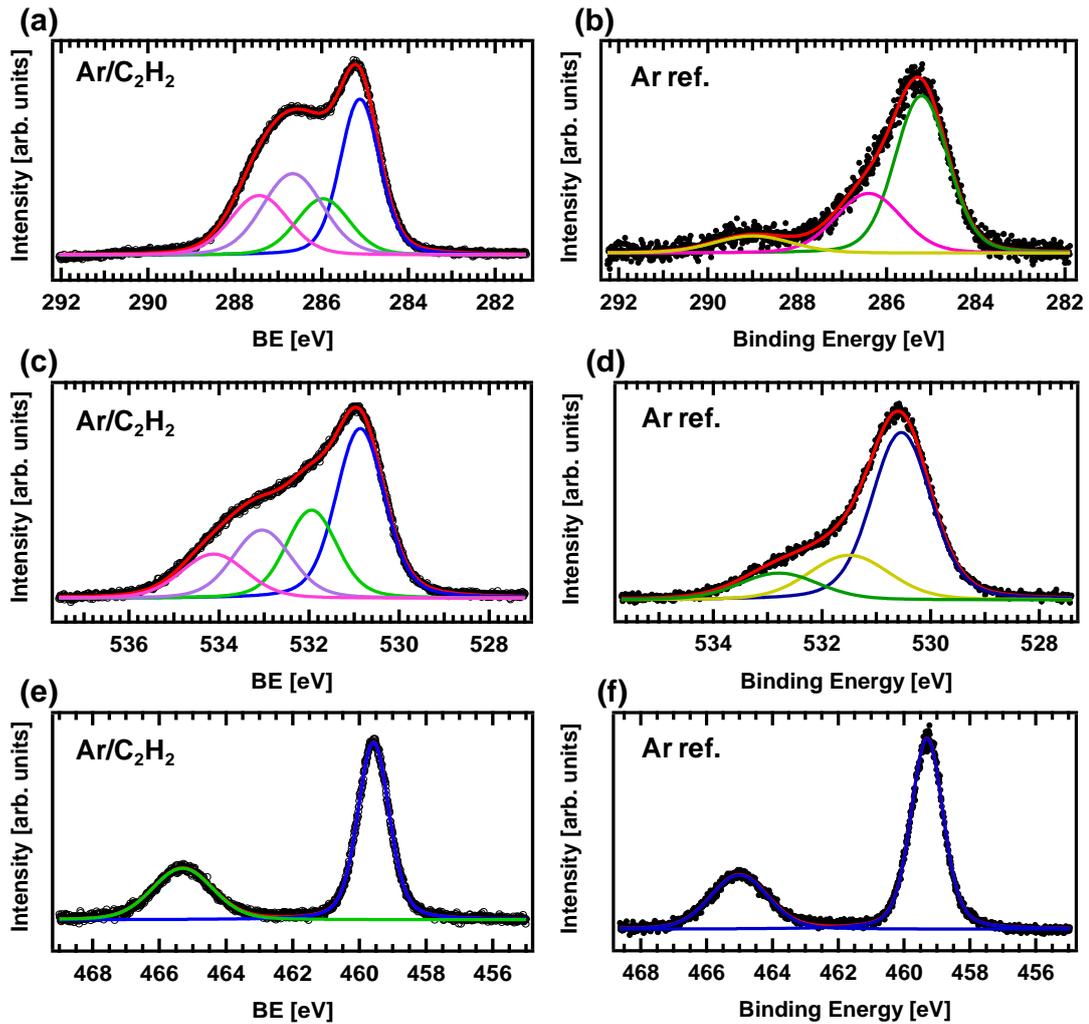

**Figure 6**. Deconvolution of the high resolution XPS peaks. (a) C1s core level peak of the one-step Ar/C$_2$H$_2$ sample. Black dots: raw data; red curve: fit. The four components identified correspond to sp$^2$, 284.8 eV (blue), another at 285.7 eV associated to C-OH (green), and two components at 286.4 eV (purple) and 287.2 eV (pink), that correspond to C-O and Ti-O-C bonds, respectively. (b) C1s core level peak of the Ar reference sample. The components identified correspond to adventitious carbon, 285.2 eV (green), and two components at 286.4 eV (pink) and 288.9 eV (yellow), that correspond to C-OH and C=O or O-C=O bonds, respectively. <u>The C-O (Ti-O-C) component does not appear</u>. (c) O1s core level peak of the Ar/C$_2$H$_2$ sample. The components identified are, one at 530.6 eV (blue) corresponding to oxygen in TiO$_2$, another at 531.7 eV (green), assigned to C-O, C-OH and Ti-O-C bonds [42, 43, 46] and finally, the contribution of C=O and O-C=O species are found at around 532.7 eV(purple) and 533.8 eV (pink), respectively [42, 47]. (d) O1s core level peak of the Ar reference sample. The components identified are, one at 530.5 eV (blue) quite strong corresponding to oxygen in TiO$_2$, another small component at 531.5 eV (yellow), assigned to C-O, C-OH or Ti-O-C bonds and finally, the contribution of C=O or SiO$_2$, species overlap at around 532.8 eV. (e, f) Ti2p core level peak of the one-step Ar/C$_2$H$_2$ and Ar reference samples. The peak can be fitted with a single doublet component. Black dots: raw data; red curve: fit; blue and green curves: doublet component. The single component located at 459.3 eV, is in agreement with the expected position for titanium in TiO$_2$ (Ti-C bond is not detected) [25].

In Table 1 below, we summarize the XPS results from the five samples analyzed. We include the binding energies of all the components identified in the high resolution C1s, O1s and Ti2p peaks (Figs. 5, S4 and 6). The main component of each peak and the contribution from Ti-O-C bond (confirmed unambiguously) are highlighted in bold.

**Table 1.** XPS analysis and main chemical states of the samples processed.

| XPS Analysis | | Ref. Sample | | | | |
|---|---|---|---|---|---|---|
| | | $Ar/C_2H_2$ two steps | $Ar/C_2H_2$ one step | Ar Ref. | $O_2/C_2H_2$ two steps | $O_2$ Ref. |
| **Peak** | **Bonds** | **Binding Energy (eV.)** | | | | |
| | **C-H[i]** | **--** | -- | **285.2** | | **285.1** |
| **C 1s** | C-sp$^2$ | **284.7** | **284.8** | -- | 285.0 | -- |
| | C-OH | 285.8 | 285.7 | -- | **285.9** | -- |
| | C-O; **(Ti-O-C)[ii]** | -- -- | 286.4- **287.2** | 286.4 -- | -- **287.2** | 286.3 -- |
| | C=O, O-C=O | 290.2 | -- | 288.9 | 289.3 | 288.9 |
| **O 1s** | TiO$_2$ | -- | **530.6** | **530.5** | 530.4 | **530.5** |
| | C-OH; C-O **(Ti-O-C)** | -- | 531.7 | 531.5 | 531.2 | 531.3 |
| | C=O, O-C=O, SiOx | -- | 532.7- 533.8 | 532.8 | **533.1[iii]** | 532.8 |
| **Ti 2p** | TiO$_2$ | -- | **459.3** | **459.3** | **459.3** | **459.3** |

(i) Adventitious carbon (CH, C-C...)
(ii) The Ti-O-C chemical state is identified as a particular case of C-O (C-O would correspond to functionalized carbon structure).
(iii) This component is contributed by several chemical bonds.

### 3.3 Optical absorption of carbon-TiO$_2$(110) structures

Figure 7 shows the UV-visible diffuse reflectance spectra of carbon-TiO$_2$ samples deposited in two steps with Ar/C$_2$H$_2$ (in black) and O$_2$/C$_2$H$_2$ (in blue) and their related photographs. For pristine TiO$_2$, both, the reference spectrum (in red) and the photograph are also included for comparison. There is no detectable difference in the reflectance-absorbance between the reference spectrum and the sample deposited with O$_2$/C$_2$H$_2$. We relate this result to the lack of oxygen vacancies or reduction [52] and the low density (5 nuclei/μm$^2$) and size of the nuclei (lateral size around 50 nm and 8-12 nm in height) for this sample. However, a remarkable absorption in the visible range for the sample grown with Ar/C$_2$H$_2$ can be seen. A grey contrast in the photography and diminished reflectance in the spectrum in Fig. 7 is clearly observed. This grey contrast is mainly related to graphitic carbon light absorption which depends on the thickness of the deposited film (few nm in this case) [53]. However, other causes of visible light absorption might be present, as generation of oxygen vacancies by reduction of the substrate that modifies the electronic band gap [54]. In the case of the Ar/C$_2$H$_2$, the C$_2$H$_2$ molecule under plasma activation can generate certain amount of H promoting superficial or even bulk reduction of the substrate in combination with the process temperature [55]. Also Ar$^+$ from the plasma can promote surface sputtering and preferential desorption of O [56]. These well-known effects [52, 57] can be ruled out with further characterization of the electrical response of the system under illumination, as we show below.

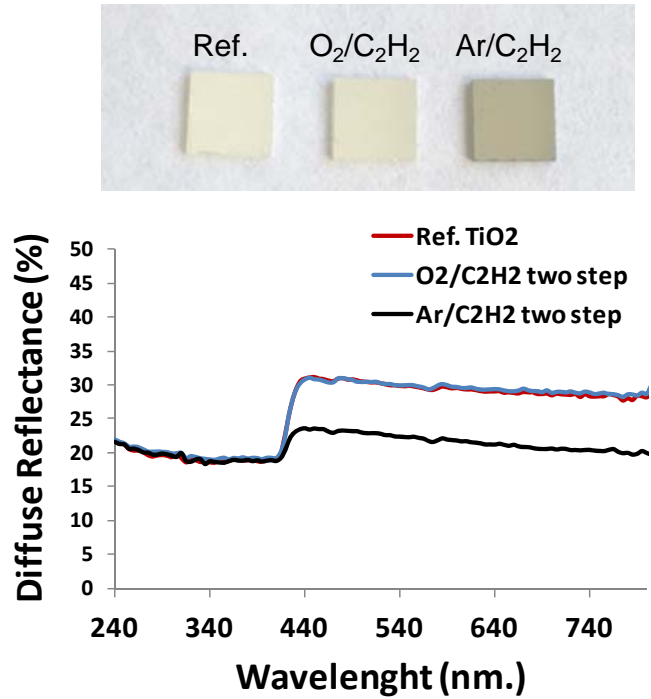

**Figure 7.** UV-visible diffuse reflectance spectra of graphene-TiO$_2$ samples grown using the two step protocol with Ar/C$_2$H$_2$ (in black), O$_2$/C$_2$H$_2$ (in blue) and their related photographs. The reference pristine TiO$_2$ substrate (in red) is also included. The absorbance values are qualitative and have a comparative purpose.

There results show that the incorporation of graphitic carbon to $TiO_2$ promotes an extended light absorption range which should eventually improve the overall performance of the material in potential applications in photocatalysis or photovoltaics [8, 18-20].

### 3.4 Photocurrent generation through carbon-$TiO_2$(110) interface

In order to test the electrical response to the light of the carbon-$TiO_2$ structure, several current *vs.* bias voltage measurements of the continuous sample deposited with $Ar/C_2H_2$ mixture in two steps have been performed. Figure 8(a) presents the set-up used to acquire the measurements [36]. With this configuration, we make contact onto the graphitic film with two carbon probes applying a voltage difference between them and obtaining the in-plane generated current both in dark conditions and after illumination. Figure 8(b) shows the current *vs.* bias voltage output characteristics of the carbon-$TiO_2$ sample in dark conditions (black line) and at different wavelengths in the UV-Vis range. Due to the high conductivity of the graphitic film, the current in dark conditions is in the order of tens of μA. The response of the sample to the visible light is negligible, as can be seen in the output (420 nm-blue line) included as an example in figure 8(b), with similar results at longer wavelengths. However, the response to the UV light is significant (300 nm-purple line). This further confirms that the chemical interaction at the interface is fairly good and the charge transfer is effective. Moreover, once the chemical contact is confirmed, the lack of excitation under visible light illumination means that the band gap of the material has not been notably modified. As the substrate reduction would decrease in a remarkable way the band gap of the material [54], this effect and its contribution to the visible light absorption can be ruled out. Accordingly, visible light absorption of the sample deposited with $Ar/C_2H_2$ in figure 7 can be mainly attributed to the graphitic layer itself.

Figure 8(c) depicts the response to the UV light at different wavelengths. The photocurrent is extracted from the total current under illumination conditions by subtracting the values in dark conditions. The response to the UV light is dependent on the wavelength and increases with the light energy at shorter wavelengths. This effect is related to the intrinsic light absorption of the $TiO_2$ in this range [6]. The detected photocurrent is in the order of μA, one order of magnitude higher than already published values of photocurrent generation, i.e. in photocatalytic $TiO_2$ nanoparticles doped with C [49], or wrapped with graphene [15]. Moreover, the lack of excitation to visible light postulates this type of carbon-TiO$_2$ junction as promising candidate in high performance UV selective photodetectors [58].

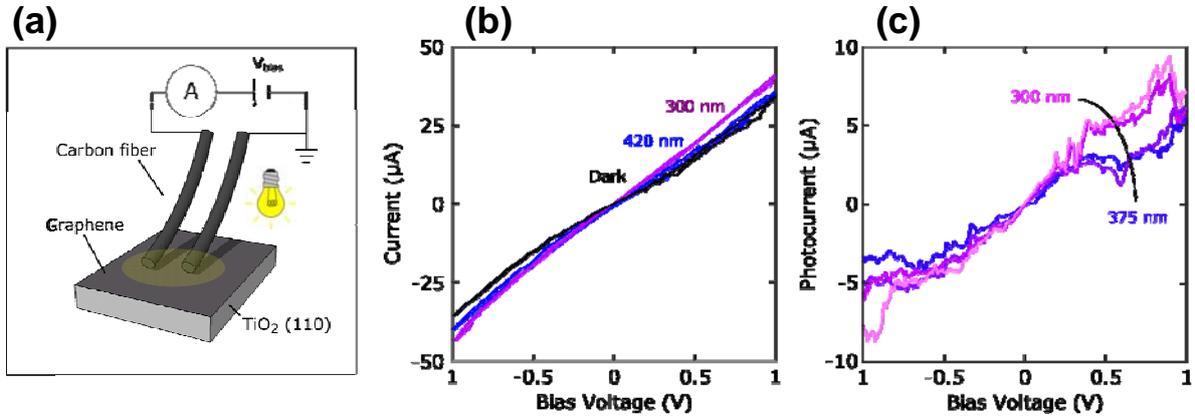

**Figure 8.** (a) Photocurrent measurement set-up. (b) Current *vs.* bias voltage output characteristics of the graphene-TiO$_2$ sample in dark conditions (black line) and in the UV-Vis range. (c) Response to the UV light at different wavelengths (from 300 nm to 375 nm). The photocurrent is extracted from the total current by subtracting the values in dark conditions (b).

## 4. Conclusions

The present study demonstrates an efficient methodology for the growth of graphenic structures on TiO$_2$. The structure, composition and growth dynamics of the deposited materials on the TiO$_2$ (110) interface can be tailored by modifying the synthesis atmosphere. On one hand, the synthesis with O$_2$/C$_2$H$_2$ results in graphene oxide nanodots/rods with controllable height and density. On the other hand the synthesis performed with Ar/C$_2$H$_2$ gases promotes the direct growth of continuous graphenic films. The measured grain size (~50 nm) and few layer thickness of the graphenic network induce a markedly low resistivity ($\rho$=6.8·10$^{-6}$ $\Omega$·m) in the films. The chemical state established between carbon and TiO$_2$ is carefully investigated by XPS and the formation of Ti-O-C chemical bonds is proposed. High photocurrent measured through carbon-TiO$_2$ interface demonstrates an efficient charge transfer that further confirms an intimate contact between both materials.

In conclusion, the synthesis methods carried out in this work led to the fabrication of structurally tailored and clean graphenic semiconductor interfaces. The reported processes are intrinsically pure, scalable and can be applied onto other semiconducting structures as nanoparticles [31].

**Acknowledgements**

We acknowledge funding from the Spanish MINECO (Grants MAT2014-54231-C4-1-P, MAT2013-47898-C2-2-R and MAT2017-85089-C2-1-R), the EU via the ERC-Synergy Program (Grant ERC-2013-SYG-610256 NANOCOSMOS), the innovation program under grant agreement No. 696656 (GrapheneCore1-Graphene-based disruptive technologies) and grant agreement No. 785219 (GrapheneCore2-Graphene-based disruptive technologies) and the Comunidad Autónoma de Madrid (CAM) MAD2D-CM Program (S2013/MIT-3007 and P2018/NMT-4367). C.S.S. acknowledges Juan de la Cierva program (IJCI-2014-19291). A.C.-G. acknowledges to grant agreement nº 755655, ECR-StG 2017 project 2D-TOPSENSE. The authors knowledge Dr. Ana Ruiz from ICMM-CSIC for technical and scientific support with XRD measurements.

**Appendix A. Supplementary data**

Supplementary data to this article can be found on line at https://doi.org/10.1016/j.apsusc.2019.144439.